\documentclass[12pt]{article}
 \newcommand{\fr}{ \frac}
 \newcommand{\beq}{\begin{equation}}
 \newcommand{\eeq}{\end{equation}}
 \newcommand{\lb}{\label}
 \newcommand{\pr}{\prime}

\begin{document}
 \begin{center}
 {\Large\bf Casimir force between  surfaces close to each other}
 \end{center}
 \vspace{5mm}
 \begin{center}
 H. Ahmedov$^1$  and I. H. Duru$^1$
 \end{center}
 1. Feza Gursey Institute,  P.O. Box 6, 81220, \c{C}engelk\"{o}y, Istanbul, Turkey
 \footnote{E-mail :  hagi@gursey.gov.tr and duru@gursey.gov.tr}. \\

 \begin{center}
 {\bf Abstract}
 \end{center}
 Casimir interactions ( due to the massless scalar field fluctuations )
 of two surfaces which are close to each  other are studied.
 After a brief general presentation of the technique, explicit calculations are performed
 for specific geometries.

 \vspace{2cm}

 \noindent
 {\Large \bf I. Introduction}

 \vspace{2mm}
 \noindent
 Experiments to observe and measure Casimir forces have so far
 been  performed with the geometrical setups involving two ( actually
 disconnected ) surfaces \cite{MOS}. The original parallel plate Casimir
 interaction is exact for infinite plane surfaces \cite{CAS}, which in
 practice means that valid for planes very close to each other.
 Effect for the parallel plane geometry were first verified in
 1958 \cite{SPAR}. Recently the experiment for this geometry
 was improved to a high precision \cite{BRES}. The Casimir experiments
 other than the above mentioned ones have been performed for a
 sphere close to  a plane configuration \cite{BROK}: which do not give rise the precise
 alignment problem of the parallel planes. Note that the
 calculation for the sphere-plane geometry gets closed to be exact
 if the radius of the sphere is small compared to the distance to
 the plane \cite{DAT}. Sphere-sphere geometry has also been studied subject
 to the similar approximation as the sphere-plane problem \cite{FEIN}. The
 interaction of two co-centric spheres has recently  been  addressed \cite{HOYE}.

 Single cavity experiments so far have not been realized \cite{HOYE},
 which we think would be very interesting: For example inserting
 the data from quantum dots ( i.e. radius $\simeq 10^{-7}$ cm )
 into the theoretical expression for the vacuum energy of a
 spherical cavity capable of confining electromagnetic field \cite{MOS},
 one gets ( in $\hbar=c=1$ units ) $ 0.5\cdot 10^6 cm^{-1} = 10 eV$
 for the Casimir energy which is of appreciable magnitude \cite{DURU}. This
 is comparable to the total energy between the parallel plates of
 the latest experiment \cite{BRES}, i.e. $E = \fr{\pi^2}{720 d^3}$(Area of
 Plates)$\simeq \fr{\pi^2}{720 (5\cdot 10^{-3})^3}\cdot (2\cdot 2)
 cm^{-1}\simeq 10^6 cm^{-1}$.

 The purpose of the present work is to study new two surfaces
 geometries. We calculate the Casimir forces resulting from the
 vacuum fluctuations of massless scalar fields between surfaces
 close to each other. For massive fields, for any realistic
 experimental setups the Casimir energies are extremely small. The
 expressions always involve a factor $e^{-\mu\bigtriangleup}$,
 where $\mu$ is the mass and $\bigtriangleup$ is the separation;
 which for electron and for nanometer distances is $e^{-2,5\cdot
 10^{10}\cdot 10^{-7}}\simeq e^{-2500}$; thus it is  practically zero.

 After a brief outline of our approach
 to the surface surface interactions in Section 2, we proceed with
 specific examples, that is co-axial cylinders, co-centric tori,
 co-centric spheres and co-axial conical  surfaces.The case of co-axial
 cylinders  my offer an experimental test in the light
 of the recent  advances in stable metal nanotubes  \cite{BAGCI}.

 \vspace{2cm}

 \noindent
 {\large \bf II. Casimir energy for the region between two boundaries
 which are close to each other}

 \vspace{2mm}
 \noindent
 We first choose the suitable spatial curvilinear coordinates
 $\eta^j$, $j=1, 2, 3$ for the geometry we deal with. The corresponding  Minkowski
 metric and the Klein-Gordon operator are then
 \beq
 ds^2=dt^2-g_{ij}d\eta^id\eta^j
 \eeq
 and ( $g\equiv \det ( g_{ij})$ )
 \beq
 \triangle = \fr{\partial^2}{\partial t^2}-\fr{1}{\sqrt{g}}
 \fr{\partial}{\partial\eta^i}g^{ij}\sqrt{g}\fr{\partial}{\partial\eta^j}.
 \eeq
 The Green function is
 \beq
  G = \sum_{\lambda_1,\lambda_2,\lambda_3} \fr{e^{i\omega (\lambda)(t-t^\prime )}}
 {2\omega(\lambda)} \Phi_{\omega(\lambda)}(\eta)  \overline{
 \Phi_{\omega(\lambda)}(\eta^\prime)},
 \eeq
 where $\Phi_{\omega(\lambda)}(\eta)$ and $\omega^2(\lambda)$ are the
 eigenfunctions and eigenvalues of the equation for the massless
 scalar field
 \beq\lb{EQ1}
 -\fr{1}{\sqrt{g}}
 \fr{\partial}{\partial\eta^i}g^{ij}\sqrt{g}\fr{\partial}{\partial\eta^j}
 \Phi_{\omega(\lambda)}(\eta) =\omega^2(\lambda)
 \Phi_{\omega(\lambda)}(\eta).
 \eeq
 ( For massive scalar field one only changes $\omega^2$ by $\omega^2
 + \mu^2$; with $\mu$ being the mass ). We assume that the above equation is
  separable in the spatial coordinates  $\eta^j$. Here $\eta$ and $\lambda$ stand for the collection of
 the coordinates $\eta^j$ and the corresponding quantum numbers $\lambda_j$ ( which are specified by the boundary conditions )
 respectively.  The  functions $\Phi_{\omega(\lambda)}(\eta)$ are normalized with
 respect to the norm
 \beq
 \|\Phi\|^2 = \int_{A} d^3\eta \sqrt{g} |\Phi (\eta)|^2,
 \eeq
 where $A$ is the domain of  the coordinates $\eta^j$.
 The vacuum energy density can be then obtained by calculating
 the coincidence limit derivatives as:
 \beq
 T = \ Reg \ [ \lim_{t,\eta^j \rightarrow
 t^\pr, \eta^{\prime j}} (\fr{\partial^2}{\partial t \partial t^\prime } + g^{ij}
 \fr{\partial^2}{\partial \eta^i\partial\eta^{\prime j}}
 ) G(\eta, \eta^\prime )].
 \eeq
 "Reg" stands for regularization. In the specific examples it means  that we have to
 subtract the  terms ( in the Plana sum formulas to be employed over the modes )
 corresponding to the  vacuum energy of the free space, the boundary energy etc.
 To calculate  the Casimir energy one needs the eigenvalues of the problem.
 The eigenvalues  $\omega^2 (\lambda)$ depend on  three quantum numbers  $\lambda_j$
 corresponding  to the degrees of freedoms in directions  $\eta^j$
 in which we assume that the equation (\ref{EQ1}) can be separated.
 We further assume that after the separation of variables the eigenvalue equations
 in coordinates $\eta^1, \eta^2$ can be trivially solved, and the corresponding quantum
 numbers $\lambda_1, \lambda_2$ are easily obtained. This assumption does not introduce
 a strong restriction. In fact many problems in the literature are of that type.
 For example when  one studies the Casimir energy inside a spherical cavity, only nontrivial
 problem is the radial equation in which one has to deal with the roots of the Bessel functions to
 impose the boundary condition \cite{MOS}.

 In this work we employ an approximation method to calculate the
 nontrivial spectral parameter $\lambda_3$, which is valid if the
 problem involves two boundaries in direction $\eta^3$, which are
 close to each other.

 After the separation, the problem in hand in $\eta_3$ can be converted into the
 Schr\"{o}dinger form
 \beq\lb{EQ2}
 [-\fr{d^2}{d(\eta^3)^2}+ W_{\lambda_1\lambda_2}(\eta_3) ]\Phi_{\lambda_3}(\eta^3)
    = E(\lambda) \Phi_{\lambda_3}(\eta^3).
 \eeq
 The form of the ''potential"  $W_{\lambda_1\lambda_2}(\eta_3)$ and the relation between $\omega^2(\lambda)$
 and $E(\lambda)$    depend on the choice of coordinate systems. The explicit examples
 are given in the following sections. The  boundary conditions we wish to impose for the type of
 geometries under investigations  are
 \beq
 \Phi_{\lambda_3}(\eta^3_0) =0, \ \ \ \Phi_{\lambda_3}(\eta^3_1) =0,
 \eeq
 where  $\eta^3_0< \eta^3_1$. In practice these boundary
 conditions require dealing with the roots of special functions
 which are  quite involved. However if the boundaries are close to
 each other, instead of (\ref{EQ2}) we can employ the simpler
 Schr\"{o}dinger equation
 \beq\lb{EQ4}
 [-\fr{d^2}{d(\eta^3)^2}+ V^0_{\lambda_1\lambda_2}(\eta_3) ]\Phi^0_{\lambda_3}(\eta^3)
    = E^0(\lambda) \Phi^0_{\lambda_3}(\eta^3)
 \eeq
 where the constant potential in the region is given by
 \beq
 V^0_{\lambda_1\lambda_2}(\eta_3) = \left\{
 \begin{array}{c}
 \infty, \ \ \ \ \eta^3=\eta^3_0, \ \eta^3=\eta^3_1 \\
 W_{\lambda_1\lambda_2}(\sqrt{\eta^3_0\eta^3_1}), \ \ \eta^3\in (\eta^3_0, \eta^3_1)   \
 \end{array}\right.
 \eeq
 The eigenvalue equation (\ref{EQ4}) has the following solutions
 \beq
 E^0(\lambda)= (\fr{\pi \lambda_3}{\triangle})^2 + W^0_{\lambda_1\lambda_2}
 \eeq
 and
 \beq
 \Phi^0_{\lambda_3}(\eta^3)= \sqrt{\fr{2}{\triangle}} \sin
 (\fr{\pi\lambda_3}{\triangle}),
 \eeq
 where $\triangle= \eta^3_1 -\eta^3_0$ and $\lambda_3 =1, 2,
 \dots$.  The system given by (\ref{EQ4}) is  a good approximation if
 the condition
 \beq\lb{EQ5}
 \max_{\eta^3\in (\eta^3_0, \eta^3_1)} |W_{\lambda_1\lambda_2}(\eta_3) -
 W^0_{\lambda_1\lambda_2}|\ll \min_{\lambda_3} |E^0(\lambda)|
 \eeq
 is satisfied.

 In the following sections we apply this approximation method to
 the specific geometries.

 \vspace{2cm}

 \noindent
 {\large \bf III. Casimir energy in the region between two close co-axial cylinders.}

 \vspace{2mm}
 \noindent
 In the  cylindrical coordinates, i.e. with the metric
 \beq
 ds^2=dt^2-dz^2-dr^2-r^2d\phi^2
 \eeq
 the eigenvalue problem we have to solve is
 \beq\lb{EV}
 -[\fr{1}{r}\fr{\partial}{\partial r} r\fr{\partial}{\partial r}
 +\fr{1}{r^2}\fr{\partial^2}{\partial \phi^2} +
 \fr{\partial^2}{\partial z^2}] \Phi = \omega^2 \Phi
 \eeq
 After solving for the trivial coordinates $z$ and $\phi$  we have
 \beq
 \Phi = \fr{e^{ipz+im\phi}}{2\pi\sqrt{r}}v_{nm}(r).
 \eeq
 Here   $v^p_{nm}(r)$ are the normalized  wavefunctions
 corresponding to the radial equation
 \beq\lb{EQ6}
 [- \fr{d^2}{dr^2} + \fr{m^2-1/4}{r^2}]v_{nm} = \mu_{nm}^2 v_{nm},
 \eeq
 with
 \beq
 \omega_{pnm} =\sqrt{ p^2 + \mu_{nm}^2}.
 \eeq
 The quantum number $n$ should be determined from the boundary
 conditions on the co-axial cylinders with the
 radii $r_0< r_1$:
 \beq\lb{B}
 v_{nm}(r_0)=0, \ \ \ \ v_{nm}(r_1)=0.
 \eeq
 The solution of (\ref{EQ6}) satisfying the boundary condition at $r_0$
 is given in terms of the Bessel functions as
 \beq
 v_{nm} (r) = \sqrt{\mu_{nm}r}
 \fr{J_m(\mu_{nm}r_0)N_m(\mu_{nm}r)-J_m(\mu_{nm}r)N_m(\mu_{nm}r_0)}{\Omega_{nm}},
 \eeq
 where  $\Omega_{nm}$ to be obtained from the  normalization
 \beq
 \int_{r_0}^{r_1}dr r|v_{nm}(r)|^2  = 1.
 \eeq
 In practice however the above integral is very difficult to
 calculate for arbitrary values of $r_0$ and $r_1$.
 The spectrum  $\mu_{nm}$ should be determined from the boundary
 condition  at $r_1$  which is quite involved  equation. However if the cylindrical
 surfaces are close to each other we can rely on the approximation method summarized
 in the previous section. Instead of the  eigenvalue problem (\ref{EQ6}) we consider the
 following  one
 \beq\lb{SE1}
 [-\fr{d^2}{dr^2}+ V(r)]v^0_{nm} = (\mu^0_{nm})^2 v^0_{nm}
 \eeq
 with the constant potential
 \beq\lb{EQ60}
 V(r) = \left\{
 \begin{array}{c}
 \infty, \ \ r=r_0, \ r=r_1 \\
 \fr{m^2-\fr{1}{4}}{r_0r_1}, \ \ r\in (r_0, r_1)  \
 \end{array}\right.
 \eeq
 The above equation is then trivially solved as
 \beq
 v^0_{nm} =
 \sqrt{\fr{2}{\triangle }}\sin (\mu^0_{nm}(r-r_0)); \ \ \ \triangle\equiv  r_1-r_0
 \eeq
 with  the spectrum
 \beq
 \mu^0_{nm} = \sqrt{\fr{\pi^2 n^2}{\triangle^2} +
 \fr{m^2-\fr{1}{4}}{r_0r_1}}; \ \ n=1, 2, 3, \dots
 \eeq
 For the present specific case the condition (\ref{EQ5})  is valid
 for $\triangle\ll r_0$. The Green function of the system is then easy to deal
 with:
 \beq
 G =\sum_{n=1}^\infty \sum_{m=-\infty}^\infty
 \int_{-\infty}^\infty dp \fr{e^{i\omega^0_{pnm}(t-t^\pr)+ip(z-z^\pr)+im(\phi -\phi^\pr)}}
 {8\pi^2\omega^{pnm}} v^0_{nm}(r)v^0_{nm}(r^\pr)
 \eeq
 where
 \beq
 \omega^0_{pnm}=\sqrt{ p^2 + (\mu_{nm}^0)^2 }.
 \eeq
 We insert the above Green function into the coincidence limit formula
 \beq
 T = \ Reg \ [ \fr{1}{2} \lim_{t,r,z,\phi \rightarrow
 t^\pr,r^\pr,z^\pr,\phi^\pr} (\partial_t\partial_{t^\pr} + \partial_r\partial_{r^\pr}+
 \partial_{z}\partial_{z^\pr}+\fr{1}{r^2}
 \partial_{\phi}\partial_{\phi^\pr}) G ],
 \eeq
 where ''Reg" stands for the usual regularization which will be
 defined explicitly. The  total vacuum  energy per unit height is
 \beq \lb{EQ7}
 E = \int_0^{2\pi} d\phi \int_{r_0}^ {r_1}rdr T =
 \fr{1}{2} \int_{-\infty}^\infty
 \fr{dp}{2\pi}\sum_{m=-\infty}^\infty  \ Reg \ [\sum_{n=1}^\infty
 \omega^0_{pnm}].
 \eeq
 To perform summations we use the Plana  formula \cite{MOS}:
 \beq\lb{EQ70}
 \sum_{n=0}^\infty F(n)=\fr{F(0)}{2} +\int_0^\infty dn F(n)+
 i\int_0^\infty dt \fr{F(it)-F(-it)}{e^{2\pi t} -1}.
 \eeq
 In the application of the above formula to the summation over the
 radial quantum number $n$, the $n=0$ term and the second term,
 corresponding to the surface singularity and free space  divergence respectively should be
 subtracted. Thus the  regularization stands for (in $n$ summation )
 \beq\lb{EQ8}
 Reg \ [\sum_{n=1}^\infty F(n)] = \sum_{n=1}^\infty F(n)+ \fr{F(0)}{2} -\int_0^\infty dn
 F(n).
 \eeq
 Going back to (\ref{EQ7}) we first perform the  sum over $m$. Note
 that since the argument of the square root in (\ref{EQ7}) is
 always positive we can replace it by the absolute value
 \begin{eqnarray}\lb{SUM}
 E & = & \fr{1}{2}\int_{-\infty}^\infty \fr{d p}{2\pi}\int_{-\infty}^\infty d m
 \ Reg \  \sum_{n=1}^\infty  \sqrt{|p^2+(\fr{\pi n}{\triangle})^2+
 \fr{m^2-1/4}{R^2}|} + \nonumber \\
 & + &  \int_{-\infty}^\infty \fr{dp}{2\pi} \ Reg \
 \sum_{n=0}^\infty \ Reg \  \sum_{m=1}^\infty
 \sqrt{|\fr{m^2}{R^2}+p^2 +(\fr{\pi n}{\triangle})^2-\fr{1}{4R^2}| } = \nonumber \\
 & = &  2\pi R E(\triangle, \fr{1}{2R} ) + E_1,
 \end{eqnarray}
 where $R^2=r_0r_1$. Here
 \beq\lb{EQ9}
 E(\triangle, \fr{1}{2R} ) = \fr{1}{2}\int_{-\infty}^\infty \fr{d^2\overrightarrow{k}}{(2\pi)^2}
 \ Reg \ [ \sum_{n=1}^\infty  \sqrt{||\overrightarrow{k}|^2+(\fr{\pi n}{\triangle})^2-
 \fr{1}{4R^2}|} \ \ ]
 \eeq
 with $\overrightarrow{k}$ being 2-dim. vector $\overrightarrow{k}= (p,\fr{m}{R})$;
 and,
 \beq
 E_1= \int_{-\infty}^\infty \fr{dp}{2\pi} \ Reg \ [
 \sum_{n=1}^\infty \ Reg \ [ \sum_{m=1}^\infty
 \sqrt{|\fr{m^2}{R^2}+p^2 +(\fr{\pi n}{\triangle})^2-\fr{1}{4R^2}| } \ \ ]].
 \eeq
 It is obvious that employment of $Reg \ \sum_{m=1}^\infty$ does
 not mean that there is an actual regularization in $m$ summation.
 It simply imply the usage of (\ref{EQ8}); for as it will be seen below that the
 first summation on the right hand side of that formula  is exactly calculable.

 To evaluate $E_1$, we first employ (\ref{EQ8}) in $n$ summation. The last term,
 i. e., the $\int_0^\infty dn$   integral term, becomes formally the same as $E$ of
 (\ref{EQ9}). Thus we can write
 \beq\lb{EQ10}
 E_1= -2\triangle E(\pi R,\fr{1}{2R}) + E_2 + E_3.
 \eeq
 The terms $E_2$ and  $E_3$ which come from the first and second terms
 of (\ref{EQ8}) are
 \beq
 E_2= \fr{1}{2} \int_{-\infty}^\infty \fr{dp}{2\pi}  \ Reg \ [ \sum_{m=1}^\infty
 \sqrt{|\fr{m^2}{R^2}+p^2 -\fr{1}{4R^2}| } \ ]
 \eeq
 and
 \beq
 E_3= \int_{-\infty}^\infty \fr{dp}{2\pi} \sum_{n=1}^\infty \ Reg \ [\sum_{m=0}^\infty
 \sqrt{|\fr{m^2}{R^2}+p^2 +(\fr{\pi n}{\triangle})^2-\fr{1}{4R^2}|} \ \ ].
 \eeq
 To evaluate $E_2$, we write (\ref{EQ8}) as the last term of the
 right hand side of  (\ref{EQ70}); then after suitable change of variables
 we arrive at
 \beq
 E_2= -\fr{1}{8\pi^2 R^2} ( \int_0^\infty \fr{x^3
 dx}{\sqrt{1+x^2}}\int_1^\infty dy \fr{\sqrt{y^2-1}}{e^{\pi xy}-1}
 + \int_0^1 \fr{x^3
 dx}{\sqrt{1-x^2}}\int_1^\infty dy \fr{\sqrt{y^2+1}}
 {e^{\pi xy}-1})
 \eeq
 We can easily estimate the upper limit of the above integrals. The
 first one is smaller than $\fr{1}{7200}$, while the second is
 smaller than $\fr{1}{6}$. Thus
 \beq
 |E_2|< \fr{1}{48\pi^2 R^2}.
 \eeq
 To evaluate $E_3$, since $\triangle\ll R$, we can first approximate it as:
 \begin{eqnarray}\lb{FF}
 E_3 & \simeq & -\int_{-\infty}^\infty \fr{d p}{\pi} \sum_{n=1}^\infty
 \int_{R\sqrt{p^2+\fr{\pi^2 n^2}{\triangle^2}}}^\infty dm
 \fr{\sqrt{\fr{m^2}{R^2} -p^2-\fr{\pi^2
 n^2}{\triangle^2}}}{e^{2\pi m}-1} \simeq \nonumber \\
 & \simeq & -\int_{-\infty}^\infty \fr{d p}{\pi} \sum_{n=1}^\infty
 \int_{R\sqrt{p^2+\fr{\pi^2 n^2}{\triangle^2}}}^\infty dm e^{-2\pi m}
 \sqrt{\fr{m^2}{R^2} -p^2-\fr{\pi^2
 n^2}{\triangle^2}}= \nonumber \\
 & = & -\fr{1}{2\pi^2}\int_{-\infty}^\infty \fr{d p}{2\pi} \sum_{n=1}^\infty
 \fr{\partial K_0(2\pi Rs\sqrt{p^2+(\fr{\pi n}{\triangle})^2})}{\partial R}.
  \end{eqnarray}
 We use  the formula  \cite{GRAD}
 \beq
 \int_0^\infty
 d\lambda K_0(x\sqrt{\lambda^2+ b^2})= \fr{\pi}{2x} e^{-xb}
 \eeq
 for the integration over $p$, then perform the summation over $n$. Finally we have
 \beq
 E_3 \simeq -\fr{1}{4\pi R \triangle} e^{-2\pi^2 \fr{R}{\triangle}}.
 \eeq
 which is negligible small.

 To calculate the main contribution (\ref{EQ9})  and the first term of of
 (\ref{EQ10}) we apply (\ref{EQ70}) and (\ref{EQ8}):
 \begin{eqnarray}\lb{EQ11}
 E(a, \mu) & = & -\int_{-\infty}^\infty
 \fr{d^2\overrightarrow{k}}{(2\pi)^2}\int_{\fr{a}{\pi}\sqrt{|k^2-\mu^2|}}^\infty
 \fr{dn}{e^{2\pi n}-1}\sqrt{\fr{\pi^2n^2}{a^2}-k^2+\mu^2} =
 \nonumber \\
 & =& - \int_\mu^\infty \fr{dk
 k}{2\pi}\int_{\fr{a}{\pi}\sqrt{k^2-\mu^2}}^\infty
 \fr{dn\sqrt{\fr{\pi^2n^2}{a^2}-k^2+\mu^2}}{e^{2\pi n}-1} \nonumber \\
 & - &  \int_0^\mu \fr{dk
 k}{2\pi}\int_{\fr{a}{\pi}\sqrt{\mu^2-k^2}}^\infty
 \fr{dn\sqrt{\fr{\pi^2n^2}{a^2}-k^2+\mu^2}}{e^{2\pi n}-1}= \nonumber \\
 & = &  -\fr{\pi^2}{1440 a^3}
 -\fr{1}{32\pi^2a^3} \int_0^{2\mu a} \fr{y^3
 dy}{e^{yx}-1} \int_1^\infty dx \sqrt{1+x^2}
 \end{eqnarray}
 For  (\ref{EQ9}) i. e.,  with  $a=\triangle$ and $\mu=\fr{1}{2R}$,
 since $\triangle\ll R$ we have
 \beq
 E(\triangle, \fr{1}{2R})\simeq -\fr{\pi^2}{1440 \triangle^3}
 -\fr{1}{192 R^2\triangle}
 \eeq
 Inspecting (\ref{EQ11}), for $a=\pi R$, $\mu =\fr{1}{2R}$ we see
 that its contribution is $10^{-2} \fr{\triangle^2}{R^2}$ times
 the second term in the above expression: thus it is also
 negligible. The final result for the Casimir energy between the
 close cylinders is then:
 \beq\lb{FINAL}
 E=-\fr{\pi^3 R}{720 \triangle^3} (
 1+\fr{15}{2\pi^2}\fr{\triangle^2}{R^2}).
 \eeq
 Note that the inclusion of the second term in the above
 expression does not contradict our approximation of (\ref{EQ60}), for the
 contribution of the first term after this approximation in the
 potential would be of the order $\fr{\triangle^3}{R^3}$. It is easy to check
 that in $\fr{R}{\triangle}\rightarrow \infty$ limit the above
 result becomes  same as parallel plate energy.

 Finally we like to remark that, for one-boundary geometries, for
 example for $D$-dimensional ball there are satisfactory
 techniques to deal with the problem involving the roots of Bessel
 functions \cite{BORDAY}. We may hope that these techniques may
 also be adopted for geometries with two-boundaries. For
 boundaries close to each other however, we can rely on the result of
 (\ref{FINAL}), for it gives the correct limit of parallel plates
 in $\fr{R}{\triangle}\rightarrow \infty$ limit.

 \vspace{2cm}

 \noindent
 {\large \bf IV.  Casimir energy in the region between two tori}

 \vspace{2mm}
 \noindent
 Problem differs from the previous one by the boundary condition.
 Instead of (\ref{B}), the solution of the  e-value equation (\ref{EV})
 should satisfy
 \beq\lb{BD}
 \Phi|_{r=r_0}=\Phi|_{r=r_1}=0, \ \ \ \Phi|_{z=0}=\Phi|_{z=L}
 \eeq
 where $L$ is the circumference of the tori. For  $\triangle\ll r_0$ we have
 \beq
 \Phi = \fr{e^{i\fr{2\pi k z}{h}+im\phi}}{2\pi\sqrt{r}} \sqrt{\fr{2}{\triangle}}
 \sin (\fr{\pi n}{\triangle}(r-r_0))
 \eeq
 and
 \beq
 \omega_{knm} =\sqrt{ (\fr{2\pi k}{L})^2  +(\fr{m}{R})^2 +(\fr{\pi
 n}{\triangle})^2-\fr{1}{4R^2} };
 \eeq
 where $k, \ m\in Z$ and $n =1, 2, 3, \dots$; and $R^2=r_0r_1$ as in the previous section.
 The total energy between the close tori is then, after employment
 of Plana formulas
 \begin{eqnarray}\lb{EMPL}
 E & = & \fr{1}{2}\sum_{m=-\infty}^\infty \ Reg [\sum_{k=-\infty}^\infty \sum_{n=1}^\infty
 \omega_{knm}]= \sum_{m=-\infty}^\infty \int_0^\infty dk \ Reg [
 \sum_{n=1}^\infty\omega_{knm}] \nonumber \\
 & + &
 \sum_{m=-\infty}^\infty\sum_{n=1}^\infty \ Reg [
 \sum_{k=1}^\infty\omega_{knm}]
 \end{eqnarray}
 Note that unlike the previous case, since the degree of freedom
 along the tori ( i. e. along z-coordinate ) is also restricted,
 we have to perform regularizations for the $k$-summation too.
 The first term on the right hand side of the above equation is
 exactly the Casimir energy for the co-axial cylinders considered in
 the previous section. Thus, we rewrite (\ref{EMPL}) as
 \beq
 E=L E_c + E^\pr
 \eeq
 In a fashion parallel to the evaluation of (\ref{FF}), taking the advantage of  $R\gg \triangle$ we can evaluate
 $ Reg \sum_{k=1}^\infty$ in $E^\pr$:
 \beq\lb{EXP}
 E^\pr = -\fr{1}{8\pi}\fr{\partial }{\partial L} \sum_{n=1}^\infty
 \sum_{m=-\infty}^\infty K_0(2L\sqrt{\fr{m^2}{R^2}+\fr{\pi^2
 n^2}{\triangle^2}})
 \eeq
 From
 \begin{eqnarray}
 \sum_{m=-\infty}^\infty K_0(2L\sqrt{\fr{m^2}{R^2}+\fr{\pi^2
 n^2}{\triangle^2}}) & = & 2\int_0^\infty dm K_0(2L\sqrt{\fr{m^2}{R^2}+\fr{\pi^2
 n^2}{\triangle^2}})+ \nonumber \\
 & + &  2\pi \int_{\fr{\pi R n}{\triangle}}^\infty
 \fr{dm J_0 ( 2L\sqrt{\fr{m^2}{R^2}+\fr{\pi^2
 n^2}{\triangle^2}})}{e^{2\pi m}-1},
 \end{eqnarray}
 using the approximation $\fr{1}{e^{2\pi m}-1}\simeq e^{-2\pi m}$
 for $m\geq \fr{\pi R n}{\triangle}$ we arrive at \cite{GRAD}
 \beq
 \sum_{m=-\infty}^\infty K_0(2L\sqrt{\fr{m^2}{R^2}+\fr{\pi^2
 n^2}{\triangle^2}}) = \pi R (\fr{e^{-2\pi \fr{L}{\triangle} n}}{2L}+
 \fr{e^{-2\pi n\fr{\sqrt{\pi^2 R^2 +L^2}}{\triangle}}}{\sqrt{\pi^2 R^2
 +L^2}}).
 \eeq
 Thus we have
 \beq\lb{useful}
 E^\pr \simeq -\fr{R}{8}\fr{\partial }{\partial L} (
 \fr{1}{2L}\fr{1}{e^{2\pi \fr{L}{\triangle}}-1} +
 \fr{1}{\sqrt{\pi^2R^2 +L^2}}
 \fr{1}{ e^{2\pi\fr{\sqrt{\pi^2R^2 +L^2}}{\triangle}}-1/}
 )
 \eeq
 Since $L> R\gg \triangle$ this contribution is negligible small
 in compared to $L E_c$.

 \vspace{2cm}

 \noindent
 {\Large \bf V. Co-axial cylindrical boxes of finite height. }

 \vspace{2mm}
 \noindent
 Instead of (\ref{BD}), the solution of the  e-value equation (\ref{EV})
 should satisfy
 \beq
 \Phi|_{r=r_0}=\Phi|_{r=r_1}=0, \ \ \ \Phi|_{z=0}=\Phi|_{z=L}=0
 \eeq
 with $L$ being the height  of the cylinders. For  $\triangle\ll r_0$ we have
 \beq
 \Phi = \fr{\sin (\fr{\pi k}{L}z) e^{im\phi}}{\pi\sqrt{r\pi}} \sqrt{\fr{2}{L \triangle}}
 \sin (\fr{\pi n}{\triangle}(r-r_0))
 \eeq
 and
 \beq
 \omega_{knm} =\sqrt{ (\fr{\pi k}{L})^2  +(\fr{m}{R})^2 +(\fr{\pi
 n}{\triangle})^2-\fr{1}{4R^2} };
 \eeq
 where $m\in Z$ and $n, k =1, 2, 3, \dots$; and $R^2=r_0r_1$ as in the previous section.
 The total energy between the close cylinders  is then
 \begin{eqnarray}\lb{TOTAL}
 E & = & \fr{1}{2} \sum_{m=-\infty}^\infty \ Reg [\sum_{k= 1}^\infty \sum_{m=-\infty}^\infty
 \omega_{knm}]= \fr{1}{2} \sum_{m=-\infty}^\infty \int_0^\infty dk \ Reg [
 \sum_{n=1}^\infty\omega_{knm}] \nonumber \\
 & - & \fr{1}{4}\sum_{m=-\infty}^\infty \ Reg [\sum_{n= 1}^\infty \omega_{0nm}] +
 \fr{1}{2}\sum_{m=-\infty}^\infty\sum_{n=1}^\infty \ Reg [
 \sum_{k=1}^\infty\omega_{knm}]
 \end{eqnarray}
 The first term on the right hand side of the above equality is
 equal to $L E_c$, where $E_c$ is the Casimir energy for the
 co-axial  cylinders geometry given by (\ref{FINAL}). The third term for
 $\fr{L}{\triangle} \geq 1$ can be explicitly calculated using its similarity with (\ref{EXP})
 of the previous section. Namely we have to multiply (\ref{EXP}) by
 $\fr{1}{2}$ and make a change $L\rightarrow 2L$. Using
 (\ref{useful}) we arrive at
 \beq
 E_2 \simeq -\fr{R}{32}\fr{\partial }{\partial L} (
 \fr{1}{4L}\fr{1}{e^{4\pi \fr{L}{\triangle}}-1} +
 \fr{1}{\sqrt{\pi^2R^2 +4L^2}}
 \fr{1}{ e^{2\pi\fr{\sqrt{\pi^2R^2 +4L^2}}{\triangle}}-1} )
 \eeq
 Since $R\gg \bigtriangleup$ we can neglect  the second term of the above
 expression. For $L>\triangle$ we have
 \beq
 E_2\simeq \fr{R}{32L} (\fr{\pi}{\triangle}+\fr{1}{4L})
 e^{-4\pi\fr{L}{\triangle}}
 \eeq
 which is small due to the exponential factor.

 Let us consider  the second term of (\ref{TOTAL}). For the sake of simplicity we omit the factor
 $-\fr{1}{4R^2}$ in the spectrum. Applying the Plana formula to the
 summation over $m$ we get
 \beq\lb{TOTAL1}
 W= \fr{\zeta (3) R}{16 \triangle^2} +W^\pr,
 \eeq
 in which
 \beq
 W^\pr\simeq \fr{1}{48 R}-\fr{\triangle \zeta (3)}{8\pi R^2}
 \eeq
 term is very small. The main contributions to the total energy than come from the first
 terms of (\ref{TOTAL}) and (\ref{TOTAL1}):
 \beq
 E=-\fr{\pi^3 RL}{720\triangle^3}+\fr{R\zeta (3)}{16\triangle^2}.
 \eeq
 Inspecting the above result we observe that the energy is positive around
 $L\leq \fr{3}{2} \triangle$ ( within our approximation ).
 Around this value of the height, the radial force $F_{rad}=-\fr{\partial E}{\partial
 \triangle}$ is repulsive. The force on the axial direction $F_{axial}=
 -\fr{\partial E}{\partial L}$ however, is repulsive for all
 values of $L$, which forces the cylinders to become of infinite
 length. When $L$ becomes longer than $\fr{3}{2}\triangle$, the
 radial force too becomes attractive.

 \vspace{2cm}
 \noindent
 {\Large \bf VI. Casimir energy between two close co-centric spheres.}

 \vspace{2mm}
 \noindent

 We employ the  spherical coordinates
 \beq
 ds^2 = dt^2-dr^2 -r^2 (d\theta^2 + \sin^2\theta d\phi^2)
 \eeq
 and  insert the solution in terms of the spherical harmonics
 \beq
 \Phi =Y^l_m(\theta,\phi)\fr{v_{ln}(r)}{r}; \ \ \ l=0, 1, 2, \dots, \ -l \le
 m\leq l
 \eeq
 into the Klein-Gordon equation (\ref{EQ1}). The resulting radial eigenvalue
 problem we have to deal is
 \beq
 [-\fr{d^2}{dr^2}+ \fr{(l+\fr{1}{2})^2}{r^2}]v_{ln}(r) = (\omega_{ln})^2
 v_{ln}(r)
 \eeq
 subject to  the boundary conditions
 \beq\
 v_{ln}(r_0)=0, \ \ \ v_{ln}(r_1)=0.
 \eeq
 Here $r_0< r_1$ are the radii of the spheres and $n$ is the
 radial quantum number to be determined by the boundary
 conditions. To satisfy the boundary conditions one has to deal
 with the roots of the radial wave function $v_{nl} (r)$ which as in the previous
 section are the Bessel functions ( with $m$ replaced by $l+1/2$). However since we are interested
 in $\triangle \equiv r_1-r_0 \ll r_0$ limit, we can proceed as we have done in the previous section.
 For the radial wave functions and the eigenvalues we obtain
 \beq
 v^0_{ln}(r) =
 \sqrt{\fr{2}{\bigtriangleup r}}\sin (\omega^0_{ln}(r-r_0)),
 \eeq
 \beq
 (\omega^0_{ln})^2 = \fr{\pi^2 n^2}{\triangle^2} +
 \fr{(l+\fr{1}{2})^2}{R^2}; \ \ \ n=1, 2, \dots.
 \eeq
 With the above approximated radial eigenfunctions and eigenvalues we can write the
 Green function as
 \beq
 G =\sum_{n=1}^\infty \sum_{l=0}^\infty \sum_{m=-l}^l
 \fr{e^{i\omega_{ln}(t-t^\pr)}}{2\omega^0_{ln}}
 v^0_{ln}(r)v^0_{ln}(r^\pr) Y^l_m
 (\theta, \phi) \overline{Y^l_m (\theta^\pr, \phi^\pr )}
 \eeq
 Integrating the  vacuum energy density
 \beq
 T = \ Reg \ [\fr{1}{2} \lim_{t,r,\theta,\phi \rightarrow
 t^\pr,r^\pr,\theta^\pr,\phi^\pr} [\partial_t\partial_{t^\pr} + \partial_r\partial_{r^\pr}+
 \fr{1}{r^2}\partial_{\theta}\partial_{\theta^\pr}+\fr{1}{r^2\sin^2\theta}
 \partial_{\phi}\partial_{\phi^\pr}] G ]
 \eeq
 over the volume between two co-centric spheres we get the total energy
 \beq\lb{EQ14}
 E =  \sum_{l=0}^\infty (l+\fr{1}{2}) \ Reg [
 \sum_{n=1}^\infty\omega^0_{ln}].
 \eeq
 Applying the Plana formula to the $n$ summation and dropping the $n=0$ term and
 the integration over $n$ we get
 \beq\lb{CAS}
 E = -\fr{2\triangle}{\pi R^2}\int_1^\infty dn F(n)
 \eeq
 where
 \beq
 F(n) = \sum_{s=\fr{1}{2}}^\infty \fr{s^3}{e^{2\fr{\triangle}{R}
 sn}-1}
 \eeq
 To use the Plana formula \cite{MOS}
 \beq
 \sum_{k=0}^\infty f(k+\fr{1}{2}) = \int_0^\infty dy f(y) -
 i\int_0^\infty dy \fr{f(iy)-f(-iy)}{1+e^{2\pi y}}
 \eeq
 we have to get rid off the poles of the function $F(n)$ at the imaginary axis
 $2\fr{\triangle}{R}ns=2i\pi m$. Thus we work with the function
 \beq
 F_\beta = \sum_{s=\fr{1}{2}}^\infty  \fr{s^3}{e^{2\fr{\triangle}{R}x(s+\beta)}-1}
 \eeq
 with $\beta > 0$. Then (\ref{CAS}) becomes
 \beq
 E= - \fr{\pi^3 R}{360 \triangle^3} + E^\pr
 \eeq
 where
 \beq
 E^\pr =\fr{1}{2\pi \triangle}\lim_{\beta\rightarrow 0}\int_0^\infty
 \fr{ds s^3}{e^{2\pi s}+1}\int_{\fr{2\triangle}{R}} dx \sqrt{x^2-(\fr{2\triangle}{R})^2}
 (\fr{1}{e^{x(\beta + is)}-1}+\fr{1}{e^{x(\beta -is)}-1})
 \eeq
 Using $\fr{2\triangle}{R}\ll 1$ we get
 \beq
 E^\pr \simeq -\fr{\pi}{288\triangle}
 \eeq
 Thus the total energy in the region between the spheres is
 \beq
 E=-\fr{\pi^3 R^2}{360\triangle^3}
 (1+\fr{5\triangle^2}{4\pi^2R^2})
 \eeq
 In $\fr{R}{\triangle}\rightarrow\infty$ it is obvious that the
 above  energy  approaches the parallel plate formula.

 \vspace{2cm}

 \noindent
 {\Large \bf VII. Casimir interactions of two close co-axial cones.}

 \vspace{2mm}
 \noindent
 The geometry we like to present in this section is two cones with common
 axis at positive $z$-direction  and appeces  at the origin. By
 close cones we mean the appex angles $\theta_1$ and $\theta_1$ are
 close to each other, that is
 \beq
 \triangle\equiv \theta_1-\theta_0 \ll \sqrt{\sin\theta_0\sin\theta_1}\equiv \Theta.
 \eeq
 In the above approximation the solutions we employe ( in
 spherical coordinates ) which vanishes at the surfaces $\theta
 =\theta_0$ and $\theta=\theta_1$ are
 \beq\lb{EQ16}
 \Phi^\omega_{nm} =\sqrt{\fr{\omega}{r}}J_{\mu_{nm}}(\omega r)
 \fr{e^{im\phi}}{\sqrt{\pi \triangle}}
 \sin (\fr{\pi n}{\triangle} ( \theta -\theta_0)),
 \eeq
 where
 \beq
 \mu_{nm}=\sqrt{(\fr{\pi n}{\triangle})^2 + (\fr{m}{\Theta})^2}
 \eeq
 The energy $\omega$ in (\ref{EQ16}) is continuous. The Green function is ( with the cut off
 factor $\beta$ )
 \beq\lb{EQ17}
 G =\sum_{n=1}^\infty \sum_{m=0}^\infty \sum_{m=-l}^l
 \fr{e^{-\beta \omega +i\omega(t-t^\pr)}}{2\omega}
 \Phi^\omega_{nm}(r,\theta,\phi)
 \overline{\Phi^\omega_{nm}(r^\pr,\theta^\pr,\phi^\pr)}.
 \eeq
 Note that in this section we employ different regularization
 method than the previous ones. The cut off method is more
 suitable for the continuous energy spectra. Inserting the
 Green function of (\ref{EQ17}) into the coincidence limit formula and
 then integrating over $\theta$ and $\phi$, we arrive at the
 vacuum energy density at $r$:
 \beq\lb{EQ18}
 E= \ \fr{1}{4\pi r} \ Reg_\beta  [ (\fr{\partial^2}{\partial r^2}
 +2\fr{\partial^2}{\partial \beta^2})
 \sum_{n=1}^\infty\sum_{m=-\infty}^\infty
 \fr{Q_{-1/2+\mu_{nm}}(1+\fr{\beta^2}{2r^2})}{r}]
 \eeq
 $Reg_\beta$ stands for the  cut off regularization, that is we pick the  finite part of
 the expression in $\beta \rightarrow 0$ limit. In
 deriving (\ref{EQ18}) we  used the formula \cite{GRAD}
 \beq
 \int_0^\infty d\omega e^{-\beta \omega} (J_\nu (\omega r))^2=
 \fr{1}{\pi r}  Q_{\nu-1/2} ( 1+ \fr{\beta^2}{2r^2}),
 \eeq
 where $Q_\nu (x)$ is Legendre function of the second kind.  We rewrite the
 expression  (\ref{EQ18}) as
 \beq
 E= \ \fr{1}{2\pi r^4} \ Reg_y[ \widehat{O} \sum_{n=1}^\infty\sum_{m=-\infty}^\infty
 Q_{-1/2+\mu_{nm}}(1+y^2)]
 \eeq
 with
 \beq
 y\equiv\fr{\beta}{\sqrt{2}r}, \ \
 \widehat{O}\equiv 1+2y\fr{\partial}{\partial y}+\fr{y^2+1}{2}\fr{\partial^2}{\partial
 y^2}.
 \eeq
 Applying the Plana formula to the summation over   $m$ we arrive
 at
 \beq
 E=E_0+E_1,
 \eeq
 where
 \beq\lb{EQ19}
 E_0= \fr{1}{\pi r^4} \ Reg_y[ \widehat{O} \sum_{n=1}^\infty \int_0^\infty dm
 Q_{-1/2+\mu_{nm}}(1+y^2)]
 \eeq
 and
 \beq
 E_1= \fr{1}{r^4} \ Reg_y[ \widehat{O}\sum_{n=1}^\infty
 \int_{\fr{\pi \Theta n}{\triangle}}^\infty  \fr{dm
 \tanh\sqrt{(\fr{m}{\Theta})^2- (\fr{\pi n}{\triangle}^2})}
 {e^{2\pi m}  -1}
 P_{-\fr{1}{2}+i\sqrt{(\fr{m}{\Theta})^2- (\fr{\pi n}{\triangle})^2}}(1+y^2)]
 \eeq
 Making use of  $\fr{\pi\Theta n}{\triangle}\gg 1$,  $\tanh x \leq 1$ and
 \beq
 Reg_y[ \widehat{O}P_{-1/2+is} (1+y^2) ]= \fr{7}{8}-\fr{s^2}{2}
 \eeq
 we get
 \beq
 |E_1| \leq \fr{1}{r^4}| \sum_{n=1}^\infty \int_{\fr{\pi \Theta n}{\triangle}}^\infty dm
  e^{-2\pi m} (\fr{7}{8}-\fr{(\fr{m}{\Theta})^2- (\fr{\pi n}{\triangle})^2}{2})|.
 \eeq
 That is
 \beq
 |E_1| \leq \fr{\Theta}{4\pi \triangle r^4}e^{-2\pi^2
 \fr{\Theta}{\triangle}}.
 \eeq
 Thus $E_1$ is  negligible small. To evaluate $E_0$, we apply the Plana formula to the
 summation over  $n$ in (\ref{EQ19}). The formula we obtain is
 \beq
 E_0 = \mathcal{E} + \fr{\Theta\triangle}{2\pi r^4}A - \fr{\Theta}{2\pi r^4}B,
 \eeq
 where
 \beq\lb{EQ20}
 \mathcal{E}=\fr{\Theta\triangle}{2\pi r^4}\int_0^\infty
 dx\int_x^\infty dy \fr{\tanh\sqrt{y^2-x^2}}{e^{2\triangle
 y}-1}(\fr{7}{4}+x^2-y^2)
 \eeq
 \beq
 A =Reg_y[ \widehat{O}\int_0^\infty  ds s Q_{-\fr{1}{2}+s}
 (1+y^2)]
 \eeq
 \beq
 B  =Reg_y[ \widehat{O}\int_0^\infty  ds  Q_{-\fr{1}{2}+s} (1+y^2)]
 \eeq
 Changing the variables $y=t$, $x^2=t^2-k^2$ (\ref{EQ20}) can be rewritten  as
 \beq\lb{EQ21}
 \mathcal{E}=\fr{\Theta\triangle}{2\pi r^4}\int_0^\infty \fr{dt
 t}{e^{2\triangle t} -1} \int_0^1 \fr{dk k}{\sqrt{1-k^2}} \tanh (kt) (
 \fr{7}{4} -k^2 t^2)
 \eeq
 Inspecting the integrals over $k$, that is the terms
 \beq
 f_1(t) = \int_0^1 \fr{dk k}{\sqrt{1-k^2}}\tanh (kt)
 \eeq
 and
 \beq
 f_2(t) = \int_0^1 \fr{dk k^3}{\sqrt{1-k^2}}\tanh (kt)
 \eeq
 we see that both approach  very fast from the value
 $f_1(0)=f_2(0)=0$ to their respective asymptotic values
 $f_1(t\rightarrow\infty ) = 1$ and $f_2(t\rightarrow\infty ) =
 0.66=\fr{2}{3}$.

 Let us treat the second term in (\ref{EQ21}) in detail. We approximate
 $f_2(t)$ as
 \beq
 f_2(t)\simeq \{ \begin{array}{c}
   at, \ \ t\in [0, b]  \\
   \fr{2}{3} \ \ t\in [b, \infty )
 \end{array}
  \eeq
 where $a$ and $b$ are both of  order 1. The second term in (\ref{EQ21})
 then becomes
 \beq\lb{EQ22}
 \mathcal{E}_2 = -\fr{\Theta\triangle}{2\pi r^4}( a\int_0^b \fr{dt t^4}{e^{2\triangle t}-1}
 +  \int_b^\infty \fr{dt t^3}{e^{2\triangle t} -1})
 \eeq
 Since $\triangle\ll 1$, we can approximate the denominator of the
 first integrand as $e^{2\triangle t} -1 \simeq 2 \triangle t$. In
 the second integral making the change of variables $ 2\triangle
 t= s$ , we can replace the lower boundary as $2b\triangle\simeq
 0$. Thus (\ref{EQ22}) becomes
 \beq
 \mathcal{E}_2 \simeq -\fr{\Theta\triangle}{2\pi r^4}( \fr{a b^4}{8\triangle} +
 \fr{1}{16\triangle^4}\int_0^\infty \fr{ds s^3}{e^s -1}) =
  -\fr{\Theta a b^4}{16\pi r^4}-\fr{\Theta\pi^3}{720
 r^4\triangle^3}.
 \eeq
 It is obvious that the first term is negligible compared to the
 second. Similar treatment shows that that the first term in (\ref{EQ21})
 gives contributions of orders $O(\triangle)$ and
 $O(\fr{1}{\triangle})$ both are small. Inspecting (\ref{EQ19}) we see that
 the second and third terms in $E_0$ are also negligible. Thus the
 final result for our Casimir energy is :
 \beq\lb{EQ23}
 E\simeq \mathcal{E}_2\simeq -\fr{\Theta\pi^3}{720 r^4\triangle^3}.
 \eeq
 Note that the above "density" is an expression obtained after
 integrating over $\Theta$ and $\phi$. If we divide (\ref{EQ23}) to the
 angular integral
 \beq
 \int_{\theta_0}^{\theta_1}\sin\theta \int_0^{2\pi}d\phi \simeq
 2\pi\Theta \triangle
 \eeq
 we obtain the energy density averaged over the angular variables:
 \beq
 E\simeq \mathcal{E}_2=-\fr{\pi^2}{1440 r^4\triangle^4}+ O
 (\triangle^{-3}).
 \eeq
 In small $\triangle$ limit the above result is in perfect agreement with the energy density
 in the region between two infinite planes with angle $\triangle$
 between them ( i. e. the wedge problem ) \cite{DOW}
 \beq
 E=-\fr{1}{1440r^4\triangle^2}
 (\fr{\pi^2}{\triangle^2}-\fr{\triangle^2}{\pi^2}).
 \eeq

 \vspace{4 cm}

 \noindent
 {\bf Acknowledgments}

 \noindent
 The authors thank Alikram Aliev for useful theoretical
 discussions; and R. O. Onofrio for bringing some of the
 experiments including theirs into our attention. One of the
 author ( I. H. Duru )acknowledges the support of Turkish Academy
 of Sciences ( TUBA ) for its support.

 \end{document}